\documentclass[a4paper, 12pt]{article}
\author{Christian Beck\footnote{School of Mathematical Sciences, Queen
Mary, University of London, London E1 4NS, UK, e-mail:
c.beck@qmul.ac.uk}\\ and \\
 Clovis Jacinto de Matos\footnote{ESA-HQ,
European Space Agency, 8-10 rue Mario Nikis, 75015 Paris, France,
e-mail: Clovis.de.Matos@esa.int}
\\
}
\title{The Dark Energy Scale in Superconductors: Innovative
Theoretical and Experimental Concepts}
\begin{document}
\maketitle \begin{abstract} We revisit the cosmological constant
problem using the viewpoint that the observed value of dark energy
density in the universe actually represents a rather natural value
arising as the geometric mean of two vacuum energy densities, one
being extremely large and the other one being extremely small. The
corresponding mean energy scale is the Planck Einstein scale
$l_{PE} =\sqrt{l_P l_E}= (\hbar G/c^3 \Lambda)^{1/4}\sim 0.037mm$,
a natural scale both for dark energy and the physics of
superconductors. We deal with the statistics of quantum
fluctuations underlying dark energy in superconductors and
consider a scale transformation from the Planck scale to the
Planck-Einstein scale which leaves the quantum physics invariant.
Our approach unifies various experimentally confirmed or
conjectured effects in superconductors into a common framework:
Cutoff of vacuum fluctuation spectra, formation of Tao balls,
anomalous gravitomagnetic fields, non-classical inertia, and time
uncertainties in radioactive superconductors. We propose several
new experiments which may further elucidate the role of the
Planck-Einstein scale in superconductors.
%We present new theoretical and
%experimental concepts to investigate the Planck Einstein scale
%$l_{PE} =\sqrt{l_P l_\Lambda}= (\hbar G/c^3 \Lambda)^{1/4}\sim
%0.037mm$ as a natural scale for quantum fluctuations and dark
%energy in superconductors. Non-classical inertial properties of
%superconductive cavities are being interpreted as being due to
%statistical fluctuations of the superconductor's spacetime
%volume. Several new experimental setups are proposed to test for
%possible enhanced gravitational effects in superconductors:
%Investigating the formation of Tao balls (as a manifestation of
%the spontaneos breaking of general covariance in
%superconductors), measuring the mean lifetime fluctuations of a
%radioactive superconductor, and comparing the time measured by
%two clocks being located inside and outside the superconducting
%cavity.

\end{abstract}

\section{Introduction}
Observational cosmology of the last decade has revealed a
composition of the universe that poses one of the greatest
challenges theoretical physics has ever faced: We are living in a
universe that exhibits accelerating expansion in (approximately)
four space-time dimensions, with a de Sitter spacetime metric with
cosmological constant $\Lambda=1.29\times10^{-52} [m^{-2}]$
\cite{Spergel,Peebles,Copeland,Padmanabhan,Padmanabhan2006}. A
small cosmological constant is equivalent to a small vacuum energy
density (dark energy density with equation of state $w=-1$) given
by
\begin{equation}
\rho_{vac \Lambda}=\frac{c^4\Lambda} {8 \pi
G}=6.21\times10^{-10}[J/m^3] \label{eq3}.
\end{equation}
%The famous cosmological constant problem is the question why this
%energy density of the vacuum is so small as compared to the
%expectations of effective field theory.

In quantum field theory, the cosmological constant counts the
degrees of freedom of the vacuum. Heuristically, we may sum the
zero-point energies of harmonic oscillators and write:
\begin{equation}
E_{vac}=\sum_i \Bigg(\frac{1}{2}\hbar \omega_i\Bigg)\label{eq4}
\end{equation}
The sum is manifestly divergent. The natural maximum cutoff to
impose on a quantum theory of gravity is the Planck frequency,
leading to the Planck energy density of the vacuum:
$\rho_{P}=E_P/l_P^3=\frac{c^{7}}{G^2
\hbar}=4.6\times10^{113}[J/m^3]$. This prescription yields an
ultraviolet enumeration of the zero-point energy that is 122
orders of magnitude above the measured value, eq.(\ref{eq3}).
This is the well-known cosmological constant problem. There is
also a natural minimum cutoff that we can impose on the sum
eq.(\ref{eq4}), which is the minimum energy,
$E_E=\sqrt{c^2\hbar^2\Lambda}$ associated with the cosmological
length scale (or Einstein length scale), $l_E:=\Lambda^{-1/2}$.
This leads to the Einstein energy density for the vacuum,
$\rho_E=c\hbar \Lambda^2=5.26\times10^{-130} [J/m^3]$, which is
121 orders of magnitude below the measured value of the dark
energy density, eq.(\ref{eq3}). Apparently, the correct value of
vacuum energy density in the universe as seen from WMAP
\cite{Spergel} observations is approximately the geometric mean of both values
(see also \cite{hsu, pad, beck04} for related work).

We may therefore ask: What is the domain of physics and
phenomenology that provides a natural scale for dark energy? In
fact, the observed dark energy scale is very similar to typical
energy scales that occur in solid state physics in the physics of
superconductors. The principal idea discussed in recent papers
\cite{Beck3, Beck4, dematos} is that this is not a random
coincidence, but that there is a deep physical reason for this
coming from a similar structure of the theory of superconductors
and that of dark energy. A possible model in this direction is
the Ginzburg-Landau theory of dark energy as developed in
\cite{Beck4}. In this model vacuum fluctuations exhibit a phase
transition from a gravitationally active to a gravitationally
inactive state at a critical frequency, similar to the
superconductive phase transition at a critical temperature. On
the experimental side, there is an interesting possibility
arising out of this approach, namely that superconductors could
be used as suitable detectors for quantum fluctuations underlying
dark energy \cite{Beck4, dematos}. In the following we will further work out this
concept. In particular, we will deal with uncertainty relations
for vacuum energy densities and space-time volumes, and deal with
the corresponding fluctuation statistics in superconductors. We
will suggest several new laboratory experiments which could be
performed to test the theoretical concepts.

We will re-investigate the cosmological constant problem by
considering an uncertainty relation between vacuum energy density
and the four-dimensional volume of the universe. We start from
the Einstein-Hilbert action and show that in this approach one
actually obtains an `inverse' cosmological constant problem: The
cosmological constant comes out {\em too small} by 120 orders of
magnitude! This naturally suggests to regard the observed dark
energy density of the universe as the geometric mean of two
values of vacuum energy, one being 120 orders of magnitude too
large and the other one being 120 orders of magnitude too small.
The corresponding mean energy scale is the Planck-Einstein scale,
corresponding to lengths of about 0.037mm, a natural scale both
for dark energy and superconductive materials.
%We also see how the correct value of the
%CC can be obtained from this uncertainty relation through the
%quantification of spacetime with Planck sized four-cells. In
%section 3 the inertial properties of superconductive cavities are
%presented in the framework of Beck and Mackey electromagnetic
%model of dark energy with a massive graviton \cite{dematos}. In
%section 4 the Einstein-Planck scale, is introduced as being the
%natural scale for dark energy in superconductors, corresponding
%to the mean geometric scale between the Planck and the Einstein
%scale.

Quite interesting phenomena may arise out of the fact that the
relevant length scale for quantum fluctuations is the
Planck-Einstein scale in superconductive materials. Basically, our
model for quantum fluctuations in the superconductor is like a
model of quantum gravity where the Planck mass $m_{P}=(\hbar
c/G)^{1/2}$ is replaced by a much smaller value, the
Planck-Einstein mass $m_{PE}=(\hbar^3 \Lambda /cG)^{1/4}$.
Formally replacing the Planck mass by much smaller values has
also been discussed in the context of extra dimensions
\cite{extra}. Our approach here does not require extra dimensions
but just a superconducting environment. We will analyse the
observed formation of large sperical clusters of superconductive
particles, so-called Tao balls \cite{tao, tao2, tao3, hirsch} in this context, a phenomenon
that is so far unexplained in the usual theory of
superconductors, but which can be understood using our current
approach. We will also suggest to measure the formal fluctuations
of space-time in the superconductor by looking at the lifetime of
radioactive superconductors, as well as by comparing two clocks,
one located inside the superconductive cavity, and the other
being located outside.

This paper is organized as follows. In section 2 we sketch the
inverse cosmological problem and describe how to obtain a natural
scale of dark energy, the Planck-Einstein scale. Properties of
this scale are summarized in section 3. In section 4 we deal with
scale transformations from the Planck scale (relevant in a
non-superconducting environment) to the Planck-Einstein scale
(relevant in a superconducting environment). We discuss several
interesting phenomena that may produce measurable effects in this
context: Cutoffs of quantum noise spectra, formation of Tao balls,
fundamental time uncertainties in radioactive superconductors and
non-classical inertia. A more detailed model for the formation of
Tao balls is discussed in section 5. Finally, in section 6 we list
some suggestions for future experiments that could further clarify
the role of dark energy and gravity in superconductors.

\section{Inverse Cosmological Constant Problem and the Uncertainty Principle}

The cosmological constant problem is the problem that the typical
value of vacuum energy density predicted by quantum field theory
is {\em too large} by a factor of $10^{120}$ as compared to
astronomical observations. Here we show that using a different
argument, one can actually get a value that is {\em too small} by
a factor of order $10^{-120}$. Hence the observed value of the
cosmological constant in nature seems not that unnatural at all,
being the geometric mean of both values.

We start from the Einstein Hilbert action
\begin{equation}
S=\int [k(R-2\Lambda)+L_M]\sqrt{-g}d^4x,
\end{equation}
where $R$ is the Ricci scalar, $g$ is the determinant of the
space-time Lorentz metric, $L_M$ is the matter Lagrangian, and the
constant $k$ is
\begin{equation}
k= \frac{c^4}{16\pi G},
\end{equation}
where $G$ is the gravitational constant. 
Einstein's field equations are invariant under complex transformations
of the space-time coordinates $x^\mu \to iy^\mu$ except for the cosmological constant
term \cite{thooft}, which is the only relevant term for our approach in the following.
The part $S_\Lambda$ of
the action corresponding to the cosmological constant $\Lambda$
can be written as a product between the vacuum energy density
$\rho_{vac\Lambda} [J/m^3]$ and the four-dimensional volume $V
[m^4]$
\begin{equation}
S_\Lambda = -\rho_{vac\Lambda} V \label{e1},
\end{equation}
where the four-dimensional volume is expressed in its covariant
form as
\begin{equation}
V=\int d^4 x \sqrt{-g}\label{e2},
\end{equation}
and the vacuum energy density is given by
\begin{equation}
\rho_{vac\Lambda}=\frac{c^4\Lambda} {8 \pi G} \label{e3}.
\end{equation}
We may now regard $\rho_{vac\Lambda}$ and $V$ occuring in
eq.~(\ref{e1}) as canonically conjugated quantities, as
previously suggested in \cite{minic, Jejjala}. In a quantum
theory of gravity, we expect that the fluctuations in one
observable are related to fluctuations in its conjugate, according
to Heisenberg's uncertainty relation. Thus
\begin{equation}
\Delta \rho_{vac \Lambda} \Delta V \sim \hbar c.\label{e4}
\end{equation}
This resembles a kind of uncertainty relation in 4-dimensional
spacetime. Substituting eq.(\ref{e3}) into eq.(\ref{e4}), we can
write eq.(\ref{e4}) as
\begin{equation}
\Delta\Lambda \Delta V \sim 8 \pi l_P^2\label{e5}
\end{equation}
where $l_P=\sqrt{G\hbar/c^3}=1.61\times10^{-35} [m]$ is the Planck
length.

Let us now assume that the universe has a finite lifetime $\tau$.
$\tau$ is bounded from below by the current age of the universe.
 From the measured Hubble constant,
$H_0=2.3\times10^{-18} [s^{-1}]$, which is of the order of the
inverse age of the universe, the four-volume of the universe can
be estimated:
\begin{equation}
\Delta V \sim V \sim  \frac{4}{3} \pi (c\tau)^4 \sim
\frac{4}{3}\pi \Big(\frac{c}{H_0}\Big)^4\sim1.2\times
10^{105}[m^4]\label{e6}
\end{equation}
%We may actually assume that the universe may not only have
%beginning but also an end, of the order of magnitude of
%$H_0^{-1}$. In this case we get a maximum spacetime value induced
%by the finite lifetime of the universe.
%Regarding the present universe as a large-scale quantum
%fluctuation with finite life time $\tau$ we may assume $ \Delta V
%\sim V $.
Substituting eq.(\ref{e6}) into eq.(\ref{e5}) a
typical order of magnitude of the cosmological constant can be
computed, regarding the present value as a quantum fluctuation:
\begin{equation}
\Delta\Lambda\sim\Lambda=5.44\times10^{-174} [1/m^2]\label{e7}
\end{equation}
This value is in total disagreement with the experimental results,
being 121 orders of magnitude {\em below} the value measured by
WMAP:
\begin{equation}
\Lambda=1.29\times10^{-52} [1/m^2]\label{e8}
\end{equation}
Apparently we get by this formal approach a different
cosmological constant problem, which we may call the {\em inverse}
(or infrared) cosmological problem. The typical order of magnitude
of the cosmological constant, as derived from this formal
approach, turns out to be 120 orders of magnitude {\em too small}
as compared to the astronomical observations. If the universe
lives much longer, the estimated value of the typical order of
magnitute of $\Lambda$ would even further decrease.

Our conclusion is that the observed value of the cosmological
constant is not so unnatural after all. It's just given by the
geometric mean of both approaches, the one starting from zeropoint
energies in quantum field theories and the one starting from an
uncertainty relation for $S_\Lambda$. Full symmetry between both
approaches is obtained if $\tau \sim H_0^{-1}$.

%This is the so call cosmological constant problem, which clearly
%shows that there is a problem with the approximate conjugate
%relation eq.(\ref{e5}), in what fixes the four-volume. The
%smallness of the measured CC relies on the largeness of the
%observed spacetime, however there is no a priori explanation for
%why the universe is big.

Let us now try to reconcile both approaches.
Following Sorkin's work \cite{17,18} in causal set theory,
fluctuations in
$\Lambda$ are inversely related to fluctuations in $V$.  The
fluctuations of relevance to us are in the number $n_{cells}$ of
Planck sized cells that fill up the four-dimensional spacetime of
the universe:
\begin{equation}
n_{cells}\sim\frac{V}{l_P^4}\Rightarrow\Delta
n_{cells}\sim\sqrt{n_{cells}}\Rightarrow\Delta V\sim\sqrt{V}
l_P^2,\label{e9}
\end{equation}
Substituting eq.(\ref{e9}) into eq.(\ref{e5}), we obtain:
\begin{equation}
\Delta\Lambda \sqrt{V}\sim 8\pi\label{e10}
\end{equation}
Substituting the value of the four-volume of the universe of
eq.(\ref{e6}) into eq.(\ref{e10}) we find a value of the
cosmological constant in agreement with the experimentally
measured value eq.(\ref{e8}):
\begin{equation}
\Delta \Lambda\sim\Lambda\sim 10^{-52}[1/m^2]\label{e11}
\end{equation}
In summary we see that the quantization of the universe's
spacetime volume with Planck sized four-dimensional cells can
solve the cosmological constant problem if we interpret the value
of cosmological constant as being due to statistical fluctuations
of the total number of cells making up this volume, according to
the uncertainty principle eq.(\ref{e5}).

\section{The Planck-Einstein Scale}

The Planck-Einstein scale corresponds to the geometric mean value
between the Planck scale, $l_P$, which determines the highest
possible energy density in the universe, and the cosmological
length scale, or Einstein scale, $l_E=\Lambda^{-1/2}$, which
determines the lowest possible energy in the universe \cite{19,20}. The
Planck-Einstein energy density is the geometric mean
$\rho_{PE}=\sqrt{\rho_P\rho_E}$ between the two energy densities,
and the Planck-Einstein length $l_{PE}=\sqrt{l_P l_E}$ is the
geometric mean of the two length scales in the universe
\cite{hsu, pad, beck04}.
In the
following table we list side by side the relevant quantities and
their numerical values:
\begin{center}
\begin{tabular}{|c|c|c|c|}
\hline & Einstein scale & Planck-Einstein Scale & Planck scale \\
\hline & $\Lambda$, $\hbar$, c, k & $\Lambda$, $\hbar$, c, k G& c, $\hbar$, k, G \\
\hline Temperature [K] &$T_E=\frac{1}{k}\sqrt{c^2\hbar^2\Lambda}$
& $T_{PE}=\sqrt{T_E T_P}$ & $T_P=\frac{1}{k}\sqrt{\frac{\hbar
c^5}{G}}$ \\
\hline &$2.95\times10^{-55}$ & $60.71$ &
$1.42\times10^{32}$ \\
\hline Time [s] &$t_E=\sqrt{\frac{1}{c^2\Lambda}}$ &
$t_{PE}=\sqrt{t_E t_P}$ & $t_P=\sqrt{\frac{\hbar
G}{c^5}}$ \\
\hline & $2.58\times10^{43}$& $1.26\times10^{-13}$ & $5.38\times10^{-44}$ \\
\hline Length [m] &$l_E=\sqrt{\frac{1}{\Lambda}}$ &
$l_{PE}=\sqrt{l_E l_P}$ & $l_P=\sqrt{\frac{\hbar
G}{c^3}}$ \\
\hline &$8.8\times10^{25}$& $3.77\times10^{-5}$ & $1.61\times10^{-35}$ \\
\hline Mass [Kg] &$m_E=\sqrt{\frac{\hbar^2 \Lambda}{c^2}}$ &
$m_{PE}=\sqrt{M_E M_P}$ & $m_P=\sqrt{\frac{\hbar
c}{G}}$ \\
\hline &$5.53\times10^{-95}$& $9.32\times10^{-39}$ & $2.17\times10^{-8}$ \\
\hline Energy [J] &$E_E=\sqrt{c^2\hbar^2\Lambda}$ &
$E_{PE}=\sqrt{E_E E_P}$ & $E_P=\sqrt{\frac{\hbar
c^5}{G}}$ \\
\hline &$4.07\times10^{-78}$& $8.38\times10^{-22}$ & $1.96\times10^{9}$ \\
\hline Energy density [$J/m^3$]
&$\rho_E=\sqrt{c^2\hbar^2\Lambda^4}$ & $\rho_{PE}=\sqrt{\rho_E
\rho_P}$ & $\rho_P=\sqrt{\frac{c^{14}}{G^4 \hbar^2}}$ \\
\hline &$5.26\times10^{-130}$& $3.73\times10^{-9}$ & $4.6\times10^{113}$ \\
\hline
\end{tabular}
\end{center}
Explicitly one has the following formulas at the Planck-Einstein
scale:
\begin{equation}
E_{PE}=kT_{PE}=\Bigg(\frac{c^7\hbar^3 \Lambda}{G}\Bigg)^{1/4}=5.25
[meV]\label{e17}
\end{equation}
\begin{equation}
m_{PE}=\frac{E_{PE}}{c^2}=\Bigg(\frac{\hbar^3
\Lambda}{cG}\Bigg)^{1/4}=9.32\times10^{-39}[Kg]\label{e18}
\end{equation}
\begin{equation}
l_{PE}=\frac{\hbar}{M_{PE}c}=\Bigg(\frac{\hbar G}{c^3
\Lambda}\Bigg)^{1/4}=0.037[mm]\label{e19}
\end{equation}
\begin{equation}
t_{PE}=\frac {l_{PE}}{c}=\Bigg(\frac{\hbar
G}{c^7\Lambda}\Bigg)^{1/4}=1.26\times10^{-13}[s]\label{e20}
\end{equation}
\begin{equation}
\rho_{PE}=\frac{E_{PE}}{l_{PE}^3}=\frac{c^4 \Lambda}{G}=104
[eV/mm^3]
\end{equation}

One readily notices that the numerical values of Planck-Einstein
quantities correspond to typical time, length or energy scales in
superconductor physics, as well as to typical energy scales for
dark energy. In previous papers it has been pointed out
\cite{Beck3, Beck4, dematos} that there could be a deeper reason
for this coincidence: It is possible to construct theories of dark
energy that bear striking similarities with the physics of
superconductors. In these theories the Planck-Einstein scale
replaces the Planck scale as a suitable cutoff for vacuum
fluctuations.

%Note that the Planck-Einstein scale$l_{PE}=0.037[mm]$ is actually
%a little bit below the cosmological dark energy length scale
%$\lambda_{DE}=(\hbar c/\rho_{DE})^{1/4}\sim0.085[mm]$ with
%$\rho_{DE}\sim3.72KeV/cm^3$ being the measured cosmological dark
%energy density as already seen above in eq.(\ref{eq1}).

\section{Scale transformation in superconductors}

Our main hypothesis in this paper, to be worked out in the
following, is that when proceeding from a normal to a
superconducting environment it makes sense to consider a scale
transformation from the Planck length to the Planck-Einstein
length, which keeps many features of the quantum physics
invariant. We will give many examples below. The scale
transformation may induce new interesting observable phenomena in
superconductors. Quantum gravity phenomena that normally happen at
the Planck scale $l_P$ only could possibly induce related
phenomena at the Planck-Einstein scale in a superconducting
environment. Due to our scale transformation, the gravitational
constant $G=\hbar c/m_P^2$ formally becomes much stronger in a
superconductor if $m_P$ is replaced by the much smaller value
$m_{PE}$. Similarly, frame dragging effects and gravitomagnetic
fields could become much stronger as well, in line with recent
experimental observations \cite{Tajmar1, Tajmar2, Tajmar3}.
% Orbits of gravitationally interacting small objects
%could be influencing each other even if their size is much
%smaller than the planetary scale, due to the changed
%gravitational interaction strength.
The vacuum energy density of vacuum fluctuations would become
much smaller as well (of the order of dark energy density). In
the following subsections we will investigate the consequences of
our scale transformation hypothesis and show that the hypothesis
is consistent with some recent experimental observations for
superconducting materials. Moreover, we will predict some new
phenomena as a consequence of the scale transformation that could
be experimentally tested.

%In the following we argue that the Planck-Einstein scale is the
%natural scale of quantum fluctuations in superconductive
%materials. The effect of the superconducting environment is
%basically that the Planck length $l_P$ is formally replaced by
%$l_{PE}$, thus zooming quantum gravity effects to a much larger
%scale. In this context we will discuss space-time fluctuations in
%radioactive superconductors. Finally, we will also argue that
%$l_{PE}$ is also the natural scale for macroscopic structure f
%ormation of spherical objects consisting of many small
%superconducting particles, so-called Tao balls \cite{tao}.
%The much larger cosmological scale
%$l_E$ is also formally replaced by $l_{PE}$, thus large scale
%gravitational phenomena have analogues at the much smaller scale
%$l_{PE}$, leading to the formation of so-called Tao balls
%\cite{tao}.

\subsection{Cutoff for vacuum fluctuations in superconductors}

As mentioned before, in quantum field theories the natural cutoff
frequency $\omega_c$ for vacuum fluctuations is given by $\hbar
\omega_c \sim m_P c^2$. This leads to the cosmological constant
problem, since the corresponding vacuum energy obtained by
integrating over all frequencies up to $\omega_c$ is much too
large. In Josephson junctions (two superconductors separated by a
thin insulator) vacuum fluctuations of the electromagnetic field
can lead to measurable noise spectra \cite{Beck0,Beck2}. This
measurability is due to the Josephson effect and the fluctuation
dissipation theorem (details in \cite{Beck3, Beck5}). However,
the maximum Josephson frequency that can be reached with a given
superconductor (and thus the cutoff frequency of measurable
vacuum fluctuations) is determined by the gap energy of the
superconductor. This gap energy of a superconductor is
proportional to $kT_c$, where $T_c$ is the critical temperature
of the superconductor under consideration (in the BCS theory, the
proportionality factor is given by 3.5). Thus measurable noise
spectra induced by vacuum fluctuations can only be measured in
superconducting Josephson junctions up to a critical value of the
order $\hbar \omega_c \sim kT_c$.

Re-interpreted in terms of our scale transformation, this means
that for superconductors the Planck scale  $m_P$ as a cutoff for
vacuum fluctuations is formally replaced by something of the order
of the Planck-Einstein scale, since the critical temperaure
$T_c=1...140 K$ of normal and high-$T_c$ superconductors is of the
same order of magnitude as the Planck-Einstein temperature
$T_{PE}=60.7K$. Hence our scale transformation hypothesis makes
sense for vacuum fluctuations and vacuum energy as observed in a
superconducting environment. The relevant scale factor is of the
order $l_{PE}/l_P\sim 10^{30}$.

\subsection{Formation of Tao balls}

When a strong electric field is applied to a mixture of
superconducting and non-superconducting particles, a remarkable
effect is observed \cite{tao, tao2, tao3, hirsch}: Millions of superconducting
micropartices of $\mu m$ size spontaneously aggregate into
spherical balls of $mm$ size. The normal particles in the mixture
do not show this behavior, only the superconducting ones. The
effect has not been explained within the conventional theory of
superconductors so far. In fact, within the conventional theory of
superconductors one expects that normal particles respond to
electrostatic fields in just the same way as superconducting ones
do. Hence the Tao effect represents an unsolved puzzle:
Superconducting and non-superconducting matter behave in a
fundamentally different way. Assuming that the superconducting and
non-superconducting particles differ in no other way, one may even
regard this effect as pointing towards a violation of the
equivalence principle, or more generally the principle of general
covariance.

In a sense the formation of Tao balls reminds us of a kind of
`planet formation' on a scale that is much smaller than the solar
system, which is possible for superconducting matter only. When
working out thus analogy, again relevant scale factors of the
order $10^{30}\sim l_{PE}/l_P\sim l_E /l_{PE}$ arise: Tao balls
have a size of order $10^{-3}m$, whereas typical planets such as
the earth have a size of order $l_E\sim10^7m$. Hence the volume of
a Tao ball is smaller than the volume of a typical planet by a
factor $10^{-30}$, and so is the mass of the Tao ball. The
possible role of gravitational forces in the formation process of
Tao balls is further discussed in section 5.
%\footnote{The
%average mass of a Tao ball, $m_{TB}\sim18,5 \mu gr$, is very close
%to the Planck mass, $m_p\sim22 \mu gr$}.
%Assume that formally Tao
%balls can interact with the rescaled gravitational constant
%$G'=\hbar c/m_{PE}^2$ arising out of our scale transformation,
%then this $G'$ is larger than the usual gravitational constant
%$G=\hbar c/m_P^2$ by a factor of order $10^{60}$. Since the mass
%of the Tao ball is smaller by a factor $10^{-30}$ as compared to
%the earth, the gravitational potential that the Tao ball is
%exhibiting on other surrounding superconducting particles is
%larger by a factor $10^{30}$. This may explain why the force
%induced by the scale transformation is strong enough to allow for
%the aggregation of other small superconducting particles. Of
%course this is only a very heuristic argumentation, a more
%detailed theory will be provided in section 5.

\subsection{Fundamental space-time uncertainty in a radioactive
superconductor}

In the following we predict a new effect for radioactive
superconductors, which arises out of the scale transformation and
which could possibly be confirmed in future experiments. We start
from an effective Planck length comparable to the Planck-Einstein
length in superconductors. We are lead to envisage that the
spacetime volume of a superconductor is made of Planck-Einstein
sized cells, $l_{PE}^4$, which will statistically fluctuate
according to eq.(\ref{e9}):
\begin{equation}
\Delta V\sim \sqrt{V} l_{PE}^2\label{211}
\end{equation}
What should we now take for the space-time volume of a
superconductor? The problem is well-defined if we consider a
superconducting material with a finite life time, a radioactive
superconductor. Let $v$ denote the volume of the superconducting
material and $\tau$ the mean life time of the radioactive
material. We then choose the 4-volume as
\begin{equation}
V=v c\tau . \label{wup1}
\end{equation}
This is similar to the approach in section 2, where we considered
a universe with a finite life time $\tau$ of order $H_0^{-1}$.
Since the 3-volume $v$ is fixed, for a superconductor in the
laboratory there can only be a time uncertainty $\Delta t$ given
by
\begin{equation}
\Delta V =v c \Delta t . \label{wup2}
\end{equation}
Putting eq.~(\ref{wup1}) and (\ref{wup2}) into (\ref{211}) we
obtain an equation for the order of magnitude of a fundamental
time uncertainty in radioactive superconductors:
\begin{equation}
\Delta t \sim \sqrt{ \frac{\tau}{cv}} l_{PE}^2
\end{equation}
To estimate some numbers, let us consider the metastable state of
a $Nb^{90m}$ superconductor with a mean life time of $\tau =34.6
[s]$. The volume of the superconductor in Tate's experiment
\cite{Tate89} (just as an example) is $v=1.28 \times 10^{-13}
[m^3]$. From this we get
\begin{equation}
\Delta t \sim 1.3 \times 10^{-6} [s].
\end{equation}
Fundamental time uncertainties of the above kind should create a
broadening of the decay energy line width $\Gamma=\hbar/\tau$:
\begin{equation}
\frac{\Delta t}{\tau} \sim \frac{ \Delta \Gamma}{\Gamma}.
\end{equation}
For the above example we get
\begin{equation}
\frac{\Delta \Gamma}{\Gamma}\sim 10^{-8}\label{e311}
\end{equation}
which is challenging to measure. The smallness of the above
number might explain why this effect has not been revealed by the
experiments of Mazaki \cite{Mazaki} on the search for a
superconducting effect on the decay of Technetium$-99m$. However
the possibility of time fluctuations in radioactive
superconductors offers a new perspective to interpret the
positive results from Olin \cite{Olin} on the influence of
superconductivity on the lifetime of Niobium$-90m$.

%Optimization of the measurement of
%this effect, requires to use clocks with error bars, $\pm\delta
%t$, inferior to the time fluctuation expected to be generated by
%the superconductive cavity enclosing one of the clocks, $\Delta
%t\sim\pm V/cv$.

\subsection{Uncertainty principle and non-classical inertia
in superconductors}

Tajmar et al. \cite{Tajmar1, Tajmar2, Tajmar3} have measured
anomalous acceleration signals around isolated accelerated
superconductors, as well as anomalous gyroscope signals around
constantly rotating superconductors. These signals can be
interpreted in terms of an anomalous gravitomagnetic field that
is about 30 orders of magnitude larger than expected from normal
gravity. We note again that $l_{PE}/l_P \sim l_{E}/l_{PE} \sim
10^{30}$, thus the effect could again stand in relation to a
scale transformation in superconductors.

A recent paper \cite{dematos} connects the anomalous
gravitomagnetic fields and non-classical inertial properties of
superconductive cavities \cite{clovis1,clovis} with the
electromagnetic model of dark energy of Beck and Mackey
\cite{Beck4}. In this approach the vacuum energy stored in a
given superconductor is given by
\begin{equation}
\rho_{vac}=\frac{\pi (\ln 3)^4}{2}\frac{k^4}{(ch)^3}T_c^4\label{e12}
\end{equation}
One defines an dimensionless parameter $\chi$ by
\begin{equation}
\chi=\frac{B_g}{\omega}=-2 \frac {g}{a}\label{e13}
\end{equation}
Here $B_g$ is the gravitomagnetic field created by a rotating
superconductor, $\omega$ the angular velocity of the rotating
superconductor, $g$ is the acceleration measured inside the
superconductive cavity and $a$ the acceleration communicated to
the superconductive cavity \cite{Tajmar03,Tajmar05}: For a cavity
made of normal matter $\chi=2$, which means that the
gravitational Larmor theorem, $B_g=2\omega$ \cite{Mashhoon}, and
the principle of general covariance, $g=-a$, are verified. For a
superconductive cavity $\chi$ turns out to be a function of the
ratio between the electromagnetic vacuum energy density contained
in the superconductor, $\rho_{vac}$ as given by eq.(\ref{e12}),
and the cosmological vacuum energy density, $\rho_{vac\Lambda}$
as given by eq.(\ref{e3}):
\begin{equation}
\chi=\frac{3}{2}\frac{\rho_{vac}}{\rho_{vac\Lambda}}\label{e14}
\end{equation}
Substituting eq.(\ref{e12}) and eq.(\ref{211}) into eq.(\ref{e4})
we obtain for the typical size of fluctuations in $\chi$ the value
\begin{equation}
\Delta \chi \sqrt{V} \sim \frac{2\pi^2}{3} l_{PE}^2\label{e22}.
\end{equation}
This can be interpreted in the sense that the inertia inside a
superconductive cavity change with respect to their classical laws
due to the superconductor's spacetime volume fluctuations. Again
the relevant length scale is the Planck-Einstein length, rather
than the Planck length.

\section{Gravitational surface tension of Tao balls}

As already mentioned in section 4.2, a Tao ball is made up of
many superconductive microparticles of size $r\sim1 \mu m$ and its
radius is of the order $a \sim 1 mm$. Therefore a Tao Ball
consists of roughly
\begin{equation}
\Bigg(\frac{a}{r}\Bigg)^3\sim10^6 microparticles. \label{ec1}
\end{equation}
Tao \emph{et al.} \cite{tao} proposed to explain the strong
cohesion into spherical balls by a new type of surface tension
$\sigma $. Also Hirsch \cite{hirsch} emphasizes that the
formation of Tao balls cannot be understood by conventional
superconductor physics.

The Tao ball surface energy $\epsilon_a$ is the product of the
Tao ball surface $4\pi a^2$ and the surface tension $\sigma$:
\begin{equation}
\epsilon_a=4 \pi a^2 \sigma \label{ec2}
\end{equation}
If there are $(a/r)^3$ separated spherical particles then the
total surface energy $\epsilon_{tot}$ is
\begin{equation}
\epsilon_{tot} =4\pi r^2 \sigma \Bigg (\frac {a}{r}\Bigg)^3\sim
\epsilon_a \frac{a}{r} \sim 100 \epsilon_a.\label{ec3}
\end{equation}
Therefore the surface energy $\epsilon_a$ of a Tao ball is just 1 $\%$ of the
total surface energy $\epsilon_{tot}$ of the separated superconductive
microparticles it consists of. This makes it plausible why
macroscopic objects form. However, the question is what type of
force creates the surface tension. It must be a force that is
strong for superconducting matter only, and it must allow for the
spherical symmetry of the objects formed.

Let us consider a homogeneous spherical body of mass $m$, uniform
density $\rho$, and radius $a$, generating a Newtonian
gravitational field.  We may formally define a gravitational
surface energy $\epsilon_g$ (also called surface pressure)
\cite{Bicak} by
\begin{equation}
\epsilon_g=G\frac{m^2}{a}
\end{equation}
where $G$ is the gravitational constant. The gravitational surface
tension $\sigma_g$ is then given by
\begin{equation}
\epsilon_g=4 \pi a^2 \sigma_g ,\label{ec4}
\end{equation}
or
\begin{equation}
\sigma_g=\frac{1}{3} G m \rho=\frac {m g_0}{4 \pi a},\label{ec5}
\end{equation}
where $g_0=Gm/a^2$ is the gravitational acceleration at the
surface of the body. Is it consistent to explain the strong
cohesion of Tao balls via a gravitational surface tension of the
type of eq.(\ref{ec5})?

To answer this question let us assume that the gravitational
acceleration $g_0$ responsible for this hypothetical surface
tension $\sigma _g$, eq.(\ref{ec5}), is generated from the
acceleration $A$ communicated to the Tao ball by the electric
field $E_0$ via its electric charge $|q|$, according to the law of
non-classical inertia in superconductors, eq.(\ref{e13}):
\begin{equation}
|g_0|=\frac{\chi}{2}|A|\label{ec6}
\end{equation}
Thus our hypothesis is that the strong applied electric
field in a Tao cell triggers the creation of gravitational fields
that are much stronger than usual, via a suitable scale transformation $G
\to G'$. From Newton's law we calculate the Tao ball acceleration
as
\begin{equation}
|A|=\frac{|q|E_0}{m_{TB}}, \label{ec7}
\end{equation}
where $m_{TB}$ is the Tao ball mass. The electric charge $|q|$
acquired by a Tao ball while bouncing between the electrodes is
given by \cite{tao}
\begin{equation}
|q|=4 \pi \epsilon_0 k_L E_0 a^2, \label{ec8}
\end{equation}
where $a$ is the radius of the Tao ball,  $k_L=1.44$ is the
dielectric constant of liquid nitrogen, and $\epsilon_0$ is the
vacuum permittivity in SI units. Substituting equations
(\ref{ec6}), (\ref{ec7}) and (\ref{ec8}) into eq.(\ref{ec5}) we
obtain the gravitational surface tension of a Tao ball as
\begin{equation}
\sigma_g=\frac{\chi}{2}k_L \epsilon_0 E_0^2 a.\label{ec9}
\end{equation}
Substituting the law defining $\chi$, eq.(\ref{e14}), into
eq.(\ref{ec9}), we see that this gravitational surface tension is
proportional to the fourth power of the critical transition
temperature $T_c$ of the superconductive material:
\begin{equation}
\sigma_g=\frac{3}{2}\frac{ (\ln 3)^4
}{4\pi}\Bigg(\frac{T_c}{T_{PE}}\Bigg)^4 k_L \epsilon_0 E_0^2
a\label{ec10}
\end{equation}
As $T_c$ increases the gravitational surface tension increases.
This is in qualitative agreement with the experimental observation: According to
\cite{tao2}, the Tao balls formed by low-$T_c$ superconductors are
weaker and easier to break than those formed by high-$T_c$ superconductors.

%Tao's experiments with NdBCO and BSCCO tend to confirm this
NdBCO high-$T_c$ superconductors have a critical temperature of
$T_c=94K$. In that case case the numerical prefactor in
eq.~(\ref{ec10}) reduces to 1, and our theory for $\sigma_g$ then
reproduces the same result that was derived in \cite{tao} for the
surface tension using a different model:
\begin{equation}
\sigma_g^{NdBCO}=k_L \epsilon_0 E_0^2 a \sim 2 \times 10^{-3}
[N/m] \label{ec11}
\end{equation}
%For the sample BSCCO ,which has a critical temperature
%$T_c^{BSCCO}=84K$, holds $E_0^2 a=0.12 kV^2/mm$, this leads Tao to
%estimate the surface tension of BSCCO balls to be:
%$\sigma^{BSCCO}=1.52 [dyn/cm]$. In that case the coefficient
%$\chi^{BSCCO}=1.27$
%, substituting this value into eq.(\ref{ec11}),
%together with the value of $E_0^2 a=0.156 kV^2/mm$ for NdBCO, we
%obtain:
%\begin{equation}
%\sigma_g^{BSCCO}=\frac{\chi^{BSCCO}}{2}\sigma_g^{NdBCO}=1.26
%[dyn/cm]\label{ec11}
%\end{equation}
%which is in good agreement with the value measured by Tao,
%$\sigma^{BSCCO}=1.52 [dyn/cm]$, indicating therefore that at least
%we cannot rule out a gravitational origin for the surface tension
%of Tao balls. It is tempting to generalize eq.(\ref{ec11}) to any
%superconductive material (SC) with critical temperature,$T_c<94
%[]$, which is the critical temperature of NdBCO used in Tao
%experiments, and would be the temperature marking the transition
%between the quantum and the classical regime of gravitation in
%solid state physics,
%\begin{equation}
%\sigma_g^{SC}=\frac{\chi^{SC}}{2}\sigma_g^{NdBCO}\label{ec12}
%\end{equation}
%Note that using the value of the gravitational surface tension of
%NdBCO expressed in $[dyn/cm]$, this expression reduces
%approximately to $\sigma_g^{SC}\sim\chi^{SC}$.

It is interesting to note that if we substitute the spacetime
volume of a Tao ball, $V=\frac{4}{3} \pi a^3 c \Delta t$, into
the space-time uncertainty relation in a superconducting
environment, eq.(\ref{e22}), we find the following expression for
the typical radius of a Tao ball:
\begin{equation}
a\sim\pi \Bigg (\frac{1}{3 \chi^2} \frac{l_{PE}^4}{c \Delta
t}\Bigg)^{1/3}\label{ec13}
\end{equation}
Assuming that the temporal length $c \Delta t$ in eq.(\ref{ec13})
is equal to the size $c\Delta t \sim 1 \mu m$ of the
microparticles forming the Tao ball, we obtain the correct order
of magnitude for the radius of Tao balls, i.e., $a\sim 0.17[mm]$
for the case of NdBCO (with $\chi^{NdBCO}=2$). This means that
the coarse-grained microstructure of Tao balls consisting of many
smaller particles of $\mu m$ size is correctly described. Tao's
experiments could be seen as the spatial counterpart of the
experiments with radioactive superconductors discussed above,
which deal with the temporal aspects of the space-time volume
fluctuations.
%This is an additional encouraging
%result, showing that gravitational physics could play a role in
%the formation process of Tao balls.

\section{Further experimental suggestions}

The scale transformation of section 4 strongly enhances
gravitational effects in a superconducting environment. At the
same time, it strongly suppresses unwanted vacuum energy. Effects
induced by the scale transformation should be measurable. A couple
of interesting laboratory tests can then be performed with
superconductors. In the following, we list a few proposals in
this direction:

\begin{enumerate}

\item Measuring the cutoff frequency of quantum noise spectra in
superconductors. This experiment is currently performed in London
and Cambridge (UK), extending previous work of Koch et al.
\cite{koch}.
\item Measuring gravitomagnetic fields
and frame dragging effects in the vicinity of rotating
superconductors. These experiments are currently performed in
Seibersdorf (Austria) and Canterbury (NZ) \cite{Tajmar3, can}. Performing
similar experiments with rotating supersolids would also be
a valuable concept \cite{Kim}.

\item
Measuring time with high precision inside and outside
superconductive cavities. Since some inertial-like effects of
rotating superconductive rings seem to propagate outside the ring
\cite{Tajmar1,Tajmar2,Tajmar3}, it would also be interesting to
probe for temporal statistical fluctuations in the neighborhood
of a superconductive material. In order to carry out this
investigation we propose to compare the measurement of time
intervals measured by two synchronized identical clocks located
inside and outside a superconductive cavity. The clocks would be
synchronized before the cavity is made superconductive, and at
constant intervals of time, the time indicated by the clocks
would be regularly compared, in order to probe for any difference
between time statistical fluctuations inside and outside the
cavity.

\item Investigating broadening phenomena of the decay energy line width in radioactive
superconductors.

\item Measuring Coriolis forces on test masses moving inside rotating
superconductive cavities. This would basically correspond to
carrying out Foucault-type pendulum experiments inside rotating
superconductive cavities.

\item Comparing the measurement of acceleration exerted on
masses inside accelerated superconductive cavities with similar
accelerations detected inside cavities made of normal materials.

\item Carrying out the famous Einstein \emph{Gedanken} elevator experiment, by
comparing the measurement of the acceleration inside a
superconductive cavity falling under the sole influence of the
earth's gravitational field with that of a cavity made of normal
materials, which would also be in a state of free fall.

\item
Repeating the small-scale tests of the gravitational inverse
square law as performed by Adelberger et al. \cite{Kapner,
Adelberger} in a superconducting environment.

\item
Investigating the formation of Tao balls \cite{tao} in more
detail. Is their formation connected with anomalous acceleration
and gyroscope signals? Place accelerometers and gyroscopes near to
the Tao cell, similar as in Tajmar's experiments \cite{Tajmar2}.

\item Investigating a rotating Tao cell. How do Tao balls form
in a rotating environment? Compare with planetary aggregation
models where $G$ is replaced by a rescaled $G'$. Similar questions could be
dealt with when a magnetic field is applied to the Tao cell \cite{tao2,
tao3}.

\item
%Assuming that a Tao ball can form a macroscopically coherent quantum object; and since
%the ratio of quadrupolar gravitational to the quadrupolar
%electromagnetic radiation power for Tao balls is 30 orders of
%magnitude above the classical dimensionless coupling for an
%electron.
%It could be interesting to investigate the possibility to use Tao
%cells as quantum transducers of gravitational radiation into
%electromagnetic radiation and vice versa,
Carrying out Hertz-like experiments with Tao balls and checking for
gravitational radiation, in line with a similar proposal of
Chiao\cite{Chiao} for electrically charged superfluid Helium
droplets.

\item Performing high-precision measurements of force fields
in superconductors
using SQUIDS and Josephson junction arrays, in line with a suggestion
of Fischer et al. \cite{fischer}

\end{enumerate}

\section{Conclusion}

In this paper we were turning the cosmological constant problem
around, to argue that there is also an inverse cosmological
constant problem where formally the cosmological constant comes
out 120 orders of magnitude {\em too small}. For the inverse
cosmological constant problem, one starts from the Einstein
Hilbert action and considers an uncertainty relation for
4-dimensional spacetime. The true value of the cosmological
constant, as observed by WMAP, is given by the geometric mean of
both approaches, the quantum field theoretical one predicting a
value 120 orders of magnitude too large and and the one starting
from the Einstein-Hilbert action, predicting a value 120 orders of
magnitude too small. This intermediate value represents the
Planck-Einstein scale, a natural scale for dark energy,
superconductors, and solid state physics in general.

We have formulated the hypothesis that in a superconducting
environment it makes sense to formally consider a scale
transformation from the Planck scale by the Planck-Einstein scale.
This scale transformation leads to a strong suppression of certain
quantum mechanical observables (such as vacuum energy density) and
strong enhancement of others (such as gravitomagnetic fields). We
have shown that these suppression and enhancement effects are
consistent with some recent experimental observations. The
formation of large superconducting spherical balls (the Tao
effect) can also be understood in this context. A fundamental
space-time uncertainty in a superconducting environment is
predicted, which can be checked by future experiments. Ultimately
non-classical inertia in superconductive cavities can be related
to fluctuations in the number of relevant
Planck-Einstein sized space-time cells in a superconductor. We believe that
it is important to perform further precision experiments with
superconductors, to fully explore the Planck-Einstein scale and to
further investigate the connection between dark energy, gravity,
and the physics of superconductors.

\section{Acknowledgement}

C.B.'s research has been supported by a Springboard fellowship of EPSRC.

\end{document}